\documentclass[12pt]{iopart}
\usepackage{graphicx}
\begin{document}
%

\title[Excitation spectrum of $Na_xCoO_2\cdot yH_2O$]
{Effect of Na content and hydration on the excitation spectrum of the
cobaltite $\bf Na_xCoO_2\cdot yH_2O$}

\author{P. Lemmens\dag\footnote[1]{To whom correspondence should be addressed
(p.lemmens@fkf.mpg.de, http://www.peter-lemmens.de)},
 V. Gnezdilov\ddag ,
 N.N. Kovaleva\dag ,
 K.Y. Choi$\|$,
 H. Sakurai\S ,
 E. Takayama-Muromachi\S ,
 K. Takada\P ,
 T. Sasaki\P ,
 F.C. Chou$^+$,
 C.T. Lin\dag\ and
 B. Keimer\dag }

\address{\dag\ Max Planck Inst. for Solid State Research,
 MPI FKF, D-70569 Stuttgart, Germany}

\address{\ddag\ B. I. Verkin Inst. for Low Temperature Physics, NASU,
 61164 Kharkov, Ukraine}

\address{$\|$\ 2. Physikalisches Inst., RWTH Aachen,
 D-52056 Aachen, Germany}

\address{\S\ Superconducting Mat. Center,
NIMS, Tsukuba, Ibaraki 305-0044, Japan}

\address{\P\ Advanced Materials Lab., NIMS, Tsukuba, Ibaraki, 305-0044, Japan}

\address{$^+$ Center for Materials Science and Engineering, MIT, Cambridge, MA 02139,
USA}

\begin{abstract}
We report on a Raman scattering study on the superconducting cobaltite $\rm
Na_xCoO_2\cdot yH_2O$ as function of Na content and hydration
(x$\approx$1/3, 3/4 and y$\approx$0, 2/3, 4/3). The observed phonon
scattering and scattering continua are analyzed in terms of lattice strain
due to the structural misfit and disorder. Hydration, due to the
intercalation of one or two $\rm H_2O$ layers, releases a part of this
strain. Our Raman data suggest a connection between disorder on the partly
occupied Na sites, the split off of the $\rm a_{1g}$ level from the other
$\rm t_{2g}$ states of $\rm Co^{4+}$ and superconductivity.

\end{abstract}


\section{Properties of $\rm Na_xCoO_2\cdot yH_2O$}
The transition metal compound $\rm Na_xCoO_2\cdot yH_2O$ (x$\approx$1/3, 3/4
and y$\approx$0, 2/3, 4/3) has an appreciable, quasi-metallic conductivity
and still shows aspects of strong electronic correlations. There is broader
consensus that these correlations play an important role for the recently
discovered superconductivity with $\rm T_c$=4.6~K \cite{takada03}, the very
large thermopower, and other anomalous transport properties
\cite{tanaka94,terasaki97,koshibae00,motohashi01,wang03a,wang03b}.

The two-dimensional structure of $\rm Na_xCoO_2$ is given by an incoherent
coupling of $\rm CoO_2$ layers with Na layers stacked along the c axis
\cite{fouassier73,jansen74,lynn03,jorgensen03}. The resulting misfit and
pronounced strain effects allow, similar to other misfit-layered oxides, for
a considerable non-stoichiometry both on the cation and the oxygen sites and
an inhomogeneity in charge distribution along the stacking direction of the
compound \cite{balsys96,karppinen03}. In $\rm Na_xCoO_2$, with
x$\approx$0.7, two partly occupied Na sites on a honeycomb lattice alternate
along the c axis with a $\rm CoO_2$ layer of edge sharing $\rm CoO_6$
octahedra \cite{balsys96}. Ionic conduction based on high Na mobility exists
at room temperature. The octahedra in the $\rm CoO_2$ layers are tilted and
have only two oxygen coordinates per layer along the c axis. These oxygen
sites contain the single Co site. In the ab plane the Co sites form a planar
triangular lattice, from geometrical point of view the perfect base for
competing magnetic interactions. In Figure~1~a) and b) two projected views
on the layered structure of $\rm Na_xCoO_2$ are shown.

\begin{figure}[t]
\centering
\includegraphics[height=10cm]{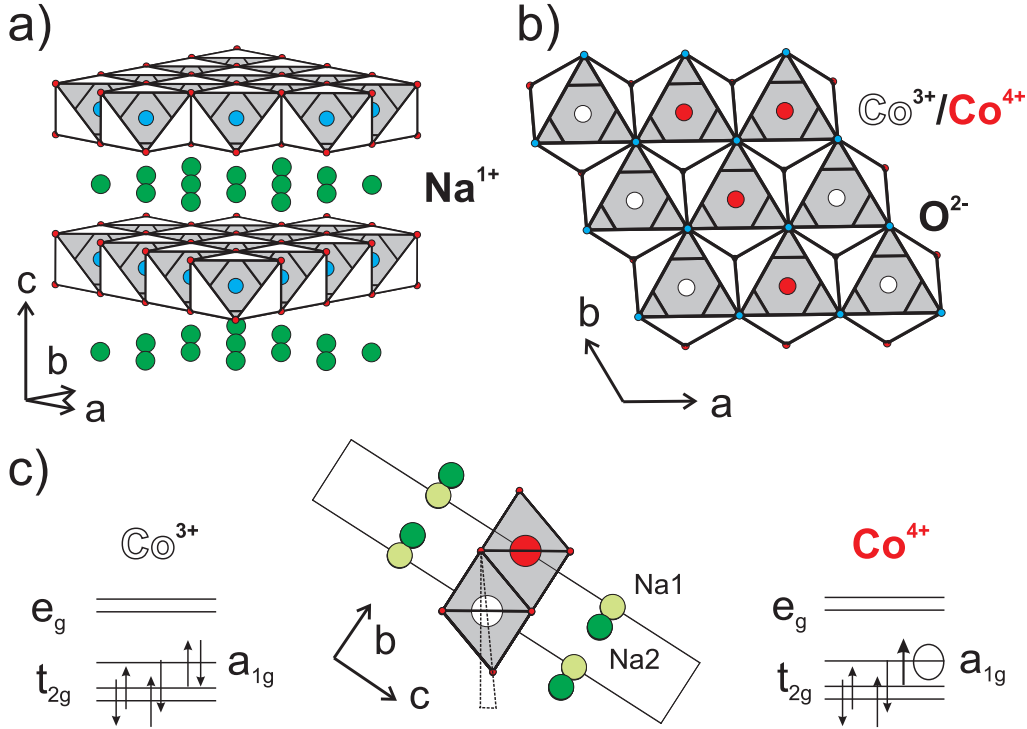}
\caption{a) and b) Two projections on a simplified structure of $\rm
NaCoO_2$ \cite{jansen74}. The Na sites in a) are only partially occupied. c)
electronic level scheme of $\rm Co^{3+/4+}$ with a sketch of two distorted
$\rm CoO_6$ octahedra \cite{balsys96}. The two sites Na1 and Na2 have an
approximate occupation 1/4 and 1/2, respectively. The line gives the unit
cell. The dashed line marks a distortion of the octahedron.} \label{sketch}
\end{figure}

Samples of $\rm Na_xCoO_2$ with a Na content 0.3$<$x$<$0.75 have been
reported. This stoichiometry range corresponds to an average Co valency
between $\rm Co^{3.7+}$ and $\rm Co^{3.25+}$. The $\rm 3d^5$ and $\rm 3d^6$
configurations of $\rm Co^{4+}$ and $\rm Co^{3+}$ occupy $\rm t_{2g}$ levels
in a low spin state with s=1/2 and s=0, respectively. Therefore, the above
mentioned triangular lattice of Co sites with s=1/2 is far from half-filled,
i.e. the spins are considerably diluted with s=0 states. In Figure~1~c) the
corresponding CEF level schemes are shown together with the local
coordination of two distorted $\rm CoO_6$ octahedra with the two Na sites,
Na1 and Na2. Relevant states at the Fermi level are given by $\rm a_{1g}$
states of $\rm 3d^5$ in $\rm Co^{4+}$ that are split off from the other $\rm
t_{2g}$ levels \cite{singh00}. This split off is due to a combination of a
trigonal distortion of the octahedra and kinetic effects \cite{koshibae03}.

The electronic state of $\rm Na_xCoO_2$ shows some analogy to strongly doped
high temperature superconductors with the additional very attractive aspect
of frustration \cite{wang03c}. Such a triangular lattice is expected to
stabilize a resonance valence bond state better than the well-studied $\rm
CuO_2$ square lattice. It also leads to three nesting vectors that could
have important implications for the superconducting order parameter
\cite{tanaka03}. Single band Hubbard models have been proposed as a
reasonable simplification taking the split off $\rm a_{1g}$ states as an
underlying basis \cite{honerkamp03,baskaran03,wang03c}. From experimental
point of view the role of electronic correlations and the minimal low-energy
model is not clear \cite{tanaka03}, as the related Mott-Hubbard insulating
phase of $\rm Na_xCoO_2$ with 1/2 doping has never been prepared or observed
and important electronic parameters could not be determined unambiguously.

The magnetic susceptibility $\rm \chi(T)$ of $\rm Na_xCoO_2$ shows a
Curie-Weiss behavior with a negative Curie constant $\rm \theta_{cw}$=-170~K
for x=0.75, while $\rm \chi(T)$ is less temperature dependent for smaller x,
i.e. $\rm \theta_{cw}$ is decreasing with increasing $\rm Co^{4+}/Co^{3+}$.
This composition dependence might be related either to spin frustration
and/or additional ferromagnetic correlations. Only in samples with highest
Na content (x=0.75) long-range magnetic ordering with ferrimagnetic or
commensurate SDW ground state is observed with a $\rm T_c$=22~K
\cite{motohashi03,sugiyama03}. In systems with smaller x long-range ordering
can only be induced by defects, as Cu substitution on the Co site
\cite{terasaki02}. Interestingly, the flat magnetic susceptibility for x=1/3
has been compared with TiOCl, a two-dimensional quantum spin system with
spin gap formation due to orbital ordering and strong spin-lattice coupled
fluctuations \cite{pickett03,seidel03,lemmens03}.

$\rm Na_xCoO_2$ has an enormous thermopower and excellent thermoelectric
figure of merit \cite{terasaki97} that are comparable to the best broad-band
semiconductors \cite{mahan97,venkatasubramanian01}. These effects origin
from the $\rm CoO_2$ layers as these layers are common structural elements
for other known thermoelectric cobaltites, as $\rm Ca_3Co_4O_9$ and $\rm
Bi_2Sr_2Co_2O_y$. The thermopower strongly depends on the ratio $\rm
Co^{4+}/Co^{3+}$. It increases with increasing x, i.e. for smaller $\rm
Co^{4+}/Co^{3+}$ and closer proximity to magnetic ordering
\cite{motohashi01}. Applying a moderate longitudinal magnetic field in the
ab plane it can be strongly depressed \cite{wang03a}. The large magnetic
field dependence rules out its origin in Fermi surface anomalies. Instead,
the combination of a large spin/electronic degeneracy and hopping on a
triangular, frustrated network has been suggested to be the origin of the
large thermopower \cite{koshibae00,wang03a}. Furthermore, the linear
temperature dependence of the Hall coefficient at high temperatures and the
linear metallic-like conductivity conductivity at low temperatures are both
proposed to be related to electronic correlations on such a triangular
lattice \cite{wang03b}.

If the Na content is strongly reduced to x=0.33-0.35 the resulting samples
are extremely hygroscopic and loosely bind $\rm H_2O$ into two intercalated
layers between the Na and the $\rm CoO_2$ layers. The result is $\rm
Na_{0.35}CoO_2\cdot 1.3H_2O$ with a nearly doubled c axis parameter
\cite{takada03}. This systems shows superconductivity with $\rm T_c$=4.6~K
and considerable experimental and theoretical evidence for an unconventional
superconducting order parameter exist \cite{tanaka03,baskaran03}. Initial
investigations show a complex defect chemistry that allows reversible
changes of the hydration level even at room temperature \cite{foo03}. On the
other side it can safely be assumed that strain related to incoherent
coupling of $\rm CoO_2$ layers to the partly occupied Na square planes is
partly released by hydration. Therefore it might also have an effect on the
electronic properties of $\rm Na_{0.35}CoO_2\cdot 1.3H_2O$. At this point
more explicit experimental information on the local site symmetries is
highly desirable.


\begin{figure}[t]
\centering
\includegraphics[height=13cm]{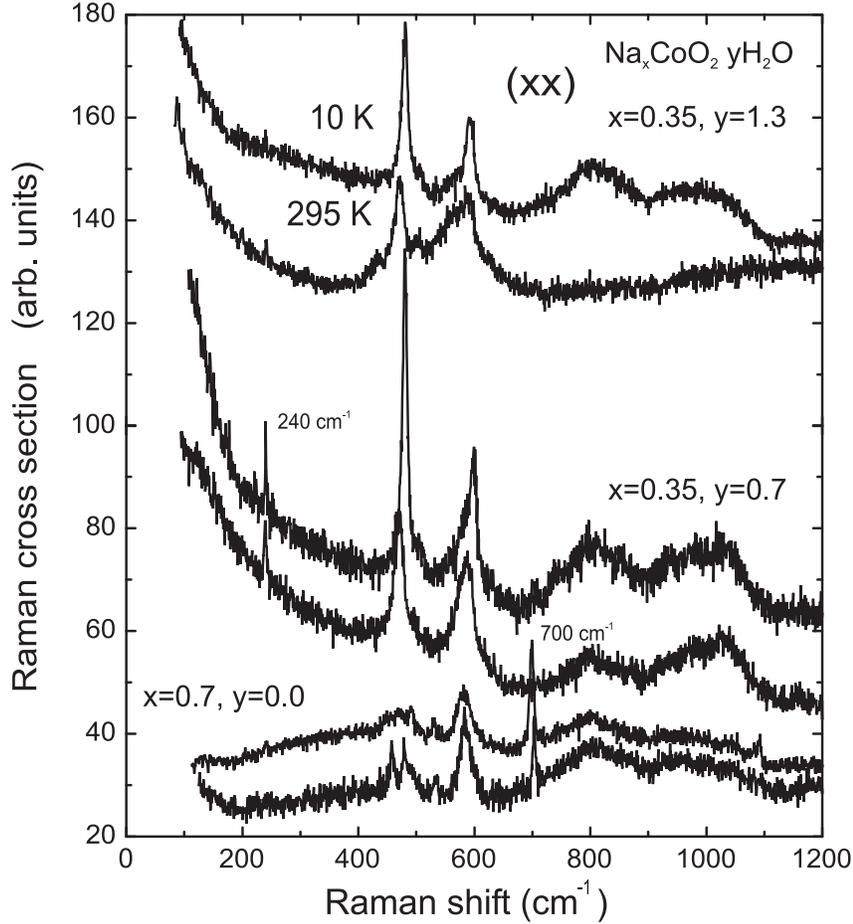}
\caption{Raman scattering intensity divided by the Bose factor of
polycrystalline $\rm Na_xCoO_2\cdot yH_2O$ as function of Na content x and
hydration y at 10K and 295~K in the upper and lower curves, respectively.
The curves are shifted for clarity.} \label{powder-sample}
\end{figure}

The existence range of the superconducting phase is very narrow
0.25$<$x$<$0.33 \cite{schaak03} or has proposed to even be a single point of
composition (x=1/3,y=4/3) \cite{jorgensen03}. The former interval motivated
theoretical investigations about possible competing charge ordered
instabilities at the phase boundaries \cite{baskaran03}. These phases would
correspond to a honeycomb and a Kagome lattice of s=1/2, respectively. Until
now no experimental evidence for such phases or the corresponding
transitions exists. It should also be noted, that the electronic structure
of $\rm Na_xCoO_2$ and the hydrated variant can be considered as Kagome-like
even without charge ordering \cite{koshibae03}.

The superconducting properties of $\rm Na_{0.35}CoO_2\cdot 1.3H_2O$ resemble
those of high temperature superconductors. The system is a type II
superconductor with the critical fields $\rm H_{c1}$=28~Oe and $\rm
H_{c2}$=61~T. The coherence lengths are $\xi$=2.3~nm and $\lambda$=570~nm
with $\kappa$=250 \cite{sakurai03}. The large critical field $\rm
H_{c2}$=61~T is exceptional if compared to the small transition temperature.
The anisotropy in transport increases with hydration from $\rm
\rho_c/\rho_{ab}$=200 to 1000. The specific heat shows a maximum at the
transition $\rm \Delta C/T_c=10.4~mJ/molK^2$ and an electronic term $\rm
\gamma=12.2~mJ/molK^2$ leading to a reduced jump of $\rm [\Delta
C/T_c]/\gamma =0.85$ considerably smaller than the BCS Value (1.43)
\cite{ueland03}. In another specific heat study a $\rm T^2$ contribution to
the specific heat is observed and attributed to line nodes of the
superconducting order parameter \cite{yang03}. Here, the other
superconducting parameters differ considerably from above: $\rm \Delta
C/T_c=7.13~mJ/molK^2$, $\rm \gamma=5.45~mJ/molK^2$ and $\rm [\Delta
C/T_c]/\gamma =1.31$. In these data a larger non-superconducting volume
fraction has been taken into account and $\rm \gamma$ corrected accordingly.



Recent NMR/NQR data show a similar lack of consistency. In one study the
Co-Knight is T-independent above $\rm T_c$ and decreases moderately for
smaller temperatures. Together with the coherence peak in NMR this would
signal a singlet order parameter with a full superconducting gap
\cite{kobayashi03}. Another experimental study highlights the negligible
temperature dependence of the Knight-shift that is taken as evidence for
spin-triplet state of superconductivity in $\rm Na_{0.35}CoO_2\cdot 1.3H_2O$
\cite{waki03}.

These and other experimental studies are strongly hampered by the easy
degradation of the sample and disorder on the Na site. The resulting small
superconducting volume fraction does not allow a sample assembly into a
vacuum setup at RT. In the present Raman scattering study special care has
been taken to avoid these problems using a setup that allows rapid cooling
down in a helium contact gas or performing the experiments at Room
temperature in a sample cell with humidified atmosphere. Thereby we searched
for possible effects of charge ordering or other instabilities and used
phonon scattering due to oxygen vibrations as a sensitive probe of the local
electronic and structural configuration.

\section{Experimental}
We have performed Raman scattering experiments on powder and single crystal
samples of $\rm Na_xCoO_2\cdot yH_2O$ as function of Na content x and
hydration y. (xx) light polarization has been investigated with the incident
and scattered electric field vector x in the ab plane of the
crystallographic structure. Powder samples were prepared as described
elsewhere \cite{takada03} and cold pressed into tablets. Single crystals of
$\rm Na_{0.7}CoO_2$ were prepared using a travelling solvent floating zone
optical furnace. In a following preparation step Na was deintercalated using
a bromine or an electrochemical preparation step \cite{chou03}. The final
hydration and/or equilibration has been performed in humidified air
\cite{schaak03}.

\section{Raman scattering results}
\begin{figure}
\centering
\includegraphics[height=10cm]{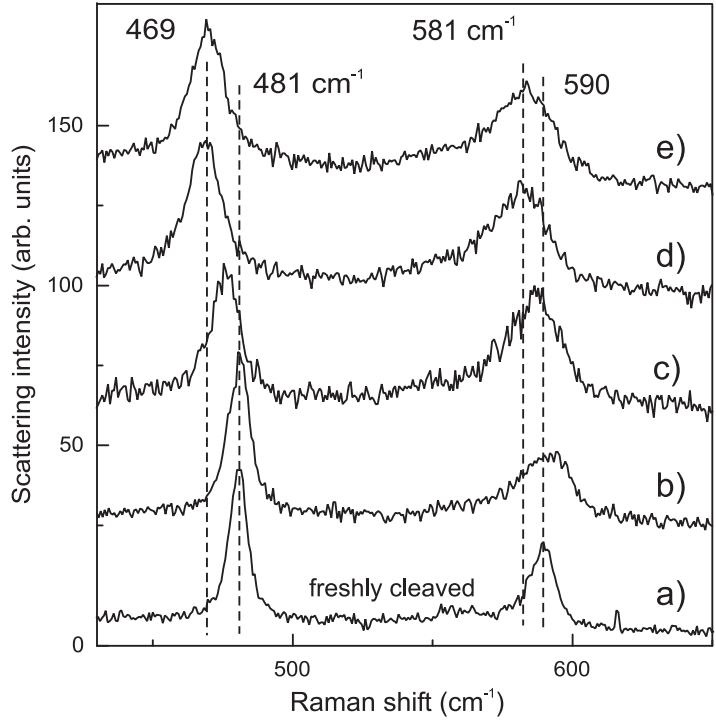}
\caption{Effect of sample environment on (xx) Raman spectra of $\rm
Na_{0.35}CoO_2\cdot 1.3H_2O$. a) freshly cleaved single crystal, b) after 2
days in humidified air, c) after 6 hours in helium exchange gas, d) after 3
hours and e) after 6 hours in vacuum. The spectra are normalized in
intensity to the mode at 470~$\rm cm^{-1}$. } \label{degradation}
\end{figure}


%

\subsection{Symmetry considerations and lattice shell model}

A symmetry analysis taking into account the $\rm P6_3/mmc$ (194) point group
for $\rm Na_{0.7}CoO_2$ \cite{balsys96} and $\rm Na_xCoO_2\cdot yH_2O$
\cite{takada03} leads to the following Raman-active modes
\begin{equation*} \rm \Gamma_{Raman} = {\bf A_{1g}} + {\bf
E_{1g}} + 3{\bf E_{2g}}, \end{equation*} and the infrared-active modes
\begin{equation*} \rm \Gamma_{IR} = 4{\bf A_{2u}} + 4{\bf E_{1u}}. \end{equation*}
Each of the $\rm E_{1g}$, $\rm E_{2g}$ or $\rm E_{1u}$ modes are doubly
degenerate.
In this analysis we skipped modes due to hydration of the samples, as no
experimental evidence for such modes exist. In the following we will also
not discuss anymore the $\rm 3E_{2g}$ modes related to Na and oxygen. These
modes would be observable in (xz) polarization \cite{iliev03}. The two Na
related modes are expected to be smeared out due to disorder and the small
occupation of the Na sites.

For the considered modes at the $\Gamma$-point the displacement symmetries
are $\rm A_{1g}$ for a displacement in the $z$ direction and $\rm E_{1g}$
for in-plane, diagonal $xy$ displacements. The $\rm A_{1g}$ and $\rm E_{1g}$
modes involve vibrations from only oxygen atoms. Due to full point-group
symmetry the Co sites do not contribute to Raman scattering.

We have performed lattice shell model calculations of the phonon frequencies
in $\rm Na_{0.74}CoO_2$ \cite{balsys96}. To accommodate for the partial
occupation of the two Na sites, Na1(0.24) and Na2(0.5), the phonon
frequencies for either a fully occupied site Na1 (or a fully occupied Na2)
site have been calculated. The frequency of the E$\rm _{1g}$ and A$\rm
_{1g}$ modes are 458~$\rm cm^{-1}$ (457~$\rm cm^{-1}$) and 586~$\rm cm^{-1}$
(574~$\rm cm^{-1}$), respectively, i.e. the higher energy A$\rm _{1g}$ mode
strongly depends on the Na site occupancy. As mentioned above the E$\rm
_{1g}$ mode is an in-plane oxygen mode with diagonal displacements while the
A$\rm _{1g}$ mode is an out-of-plane mode. This explains the strong
sensitivity of this excitation to the Na content in the layers that divide
the oxygen octahedra in c axis direction. In the following we will show that
the pronounced frequency dependence of the A$\rm _{1g}$ mode can be used as
a very susceptible sensor of Na distribution/ordering.


\subsection{Raman scattering on powder samples}
\begin{figure}
\begin{center}
\includegraphics[height=9cm]{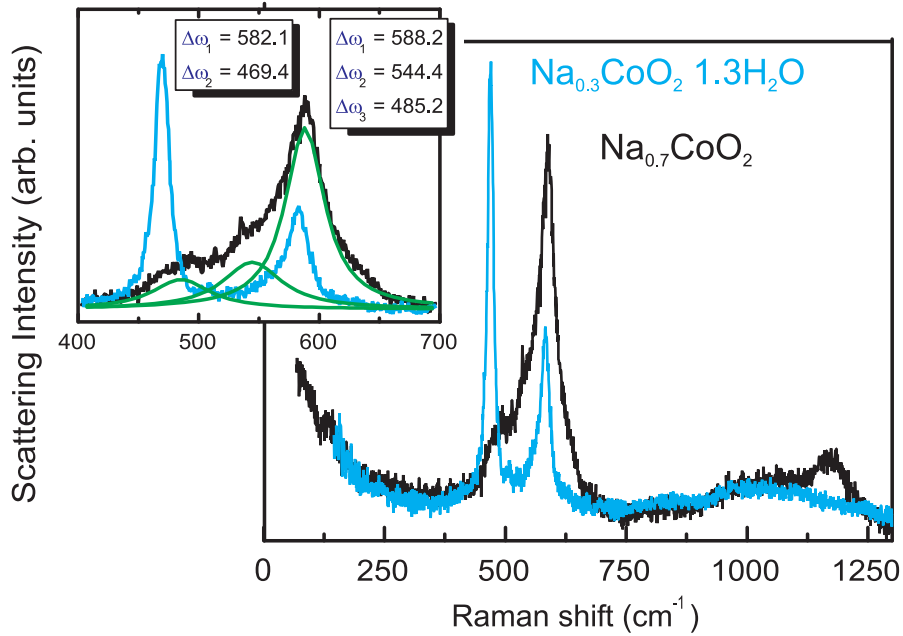}
\caption{Comparison of single crystal spectra of superconducting and
nonsuperconducting samples $\rm Na_xCoO_2\cdot yH_2O$ in (xx) polarization
at room temperature. The inset displays spectra on an enlarged scale
together with fits. For x,y=0.7/0.0 three individual modes are resolved that
overlap and form a broad band. } \label{comparison}
\end{center}
\end{figure}

Raman spectra of polycrystalline samples as given in Figure~2 show two sharp
modes at 480~$\rm cm^{-1}$ and 598~$\rm cm^{-1}$. These modes correspond to
the E$\rm _{1g}$ and the A$\rm _{1g}$ eigenmodes, respectively. The E$\rm
_{1g}$ mode hardens due to anharmonicity with decreasing temperature. The
linewidth of the A$\rm _{1g}$ mode is substantially broadened at higher
temperature. This must be related to pronounced Na diffusion and disorder.
At higher energies two very broad maxima exist at 800 and 1000~$\rm
cm^{-1}$. These signals are understood as two-phonon scattering. It is
noteworthy, that the temperature dependence of this signal is most
pronounced for samples that show superconductivity.

Reducing the hydration to zero, drastic effects happen at lower frequency. A
broadened, quasi-elastic scattering is suppressed. This continuum might be
related to quasi-diffusive excitations of $\rm H_2O$ molecules. A second
possibility are local paramagnetic fluctuations of the spin system. Such
paramagnon scattering has been discussed for high temperature
superconductors or other low-dimensional quantum spin systems
\cite{lemmens-rev}. The continuum observed in $\rm Na_xCoO_2\cdot yH_2O$
resembles to early observations in $\rm MnF_3$ \cite{richards74} and would
correspond to an exchange coupling constant of approximately 130~K. Its
highest intensity is observed for y=0.7, i.e. in the sample with one $\rm
H_2O$ layer. Other sharp modes that do not show a consistent stoichiometry
dependence, e.g. at 240 and 700~$\rm cm^{-1}$, are attributed to defect
modes.


\subsection{Raman scattering on single crystals}
Raman scattering experiments on single crystal surfaces do in general
provide a better and more consistent view on the excitation spectrum of a
compound. In the case of $\rm Na_xCoO_2\cdot yH_2O$, however, due to the
complex defect chemistry given by easy hydration loss, high mobility of Na
and its hydroxide formation special precautions have to be taken.

To investigate such effects results of a helium gas and vacuum exposure
study are shown in Figure~3. A similar study on $\rm Na_{0.7}CoO_2$ can be
found in Ref.~\cite{iliev03}. The initial state of our $\rm
Na_{0.35}CoO_2\cdot 1.3H_2O$ single crystal (curve a) has been prepared by a
freshly cleaved surface. The sample has then been stored and investigated in
humid air for two days (b) and then mounted into a cryostat with Helium
contact gas. Consecutive spectra have been taken after a storage at RT in
dry Helium gas (c) and later under vacuum (d and e). These steps were
interrupted by a cooling down to 200~K or below with cooling rates of 2-4
K/min.

It is evident from these data that the intrinsic phonon linewidth of the
fully hydrated, single crystalline $\rm Na_xCoO_2\cdot yH_2O$ is very small.
However, surface properties of the crystals show a degradation even under
humidified air. Therefore, the linewidth of the c-axis $\rm A_{1g}$ mode
broadens. The following reduction of hydration in dry helium gas leads to a
moderate decrease of the in-plane $\rm E_{1g}$ mode by about 3~$\rm
cm^{-1}$. Most drastic is the effect of a vacuum treatment. The observed
shift by 9~$\rm cm^{-1}$ is exceptionally large and as the related mode is
an in-plane oxygen vibration can not fully be accounted by the loss of
hydration. Therefore we propose a loss of oxygen as the most probable origin
of the phonon frequency shift in vacuum.

To figure out the effect of hydration on the "parent compound" (xx)
polarized Raman spectra of single crystals at T=295~K are given in Figure~4.
In addition to two-phonon scattering with a maximum at 1170~$\rm cm^{-1}$
and modes in the anti-Stokes regime at -470 and -590~$\rm cm^{-1}$ for x=0.3
and 0.7, respectively, there are remarkable changes of the phonon modes in
the frequency regime from 450-650~$\rm cm^{-1}$. $\rm Na_{0.7}CoO_2$ shows a
strongly broadened band-like scattering intensity that has a maximum at
588~$\rm cm^{-1}$, very close to the frequency calculated in a lattice shell
model with fully occupied Na1 site. In $\rm Na_{0.3}CoO_2\cdot 1.3H_2O$, in
stead, two modes are observed at 582 and 469~$\rm cm^{-1}$. The inset shows
a fit using Lorentzians to these intensities. Three modes with different
frequencies can be fitted to the broad band of $\rm Na_{0.7}CoO_2$.

We interpret the broad band of scattering in $\rm Na_{0.7}CoO_2$ as due to a
considerable strain and possible sublattice formation of the not fully
occupied Na1 and Na2 sites. With decreasing temperature a part of this
broadening is reduced, as e.g. also seen in the experiments on
polycrystalline samples in Figure~2. Ordering processes have been evidenced
from $\rm ^{23}Na-NMR$ in the temperature regime 250-300~K and attributed to
a charge disproportionation of the effective $\rm Co^{3.4+}$ into $\rm
Co^{3+}$ $\rm Co^{4+}$ \cite{gavilano03}. The sensitivity of the c-axis $\rm
A_{1g}$ mode frequency on Na site occupation and the high mobility of Na at
room temperatures, however, do not support this interpretation.

In the superconducting, hydrated system the phonon spectrum is not broadened
although the Na sublattice is even more diluted. Therefore we conclude, that
the intercalated and well-ordered $\rm H_2O$ layers \cite{jorgensen03}
shield the related disorder. As the observed phonon frequencies correspond
to in-plane and out-of-plane eigenmodes of the $\rm CoO_6$ octahedra and
their trigonal distortion determines the above discussed Co $\rm a_{1g}$
level energy, an effect of disorder on this level and the density of states
at $\rm E_{F}$ \cite{singh00} is unquestionable. The observed broadenings
and frequency shifts therefore suggest that the narrow superconducting phase
space as function of x \cite{schaak03} and the reduced superconducting
volume observed in specific heat measurements \cite{ueland03} are rather
connected with Na disorder than with the proposed charge ordering
instabilities that compete with superconductivity \cite{baskaran03}. Further
well-controlled studies as function of stoichiometry, disorder and
temperature are underway.

\section{Conclusions}
Our Raman scattering investigations on the superconducting cobaltites $\rm
Na_xCoO_2\cdot yH_2O$ as function of Na content and hydration show
pronounced effects related to partial occupation and disorder on the Na
site. In contrast, no evidence for charge ordering has been found. The
strong sensitivity of the transition temperature on local lattice
distortions is proposed to origin from a local modulation of the $\rm
a_{1g}$ split off from the remaining Co $\rm t_{2g}$ levels. These effects
support models of superconductivity based on electronic correlations.

\subsection*{Acknowledgments}
The authors acknowledge fruitful discussions with R. Kremer, I. Nekrasov,
V.I. Anisimov, R. Zeyer, P. Horsch, and G. Khaliullin. This work was
supported by DFG SPP1073, NATO PST.CLG.9777766, INTAS 01-278, CREST of JST
(Japanese Science and Technology Corporation) and the MRSEC Program of the
National Science Foundation under award number DMR 02-13282.

\end{document}